\documentclass[12pt,twoside]{article}
\usepackage{JCST}
\usepackage{graphicx}
\usepackage{amsmath,amssymb}
\begin{document}
\setcounter{page}{1}

\setpageinformation
{Zi-Ke Zhang, Tao Zhou, Yi-Cheng Zhang. Tag-Aware Recommender Systems: A State-of-the-art Survey}
{}{ }{}{Mon.}{Year}
{Zi-Ke Zhang {\it et al}.: Tag-Aware Recommender Systems: A State-of-the-art Survey}

\begin{CJK}{GBK}{song}

\title{Tag-Aware Recommender Systems: A State-of-the-art Survey}

\author{Zi-Ke Zhang$^{1,2,3,*}$, Tao Zhou$^{2,4}$, Yi-Cheng Zhang$^{1,2,3,*}$}
  \address{1}{ Institute for Information Economy, Hangzhou Normal University - Hangzhou 310036, P. R. China}
  \address{2}{ Web Sciences Center, University of Electronic Science and Technology of China - Chengdu 610054, P. R. China}
  \address{3}{ Department of Physics, University of Fribourg - Chemin du Mus\'{e}e 1700 Fribourg, Switzerland}
  \address{4}{ Department of Modern Physics, University of Science and Technology of China - Hefei 230026, P. R. China}

\vskip 1mm
\noindent E-mail: \{zhangzike@gmail.com; zhutou@ustc.edu;  yi-cheng.zhang@unifr.ch\}


\begin{abstract}
In the past decade, Social Tagging Systems have attracted increasing attention from both physical and computer science communities. Besides the underlying structure and dynamics of tagging systems, many efforts have been addressed to unify tagging information to reveal user behaviors and preferences, extract the latent semantic relations among items, make recommendations, and so on. Specifically, this article summarizes recent progress about tag-aware recommender systems, emphasizing on the contributions from three mainstream perspectives and
approaches: network-based methods, tensor-based methods, and the topic-based methods. Finally, we outline some other tag-related works and future challenges of tag-aware recommendation algorithms.\end{abstract}


\begin{multicols}{2}
\normalsize
\section{Introduction}
The last few years have witnessed an explosion of information that
the exponential growth of the Internet \cite{ZhangGQ200801} and World Wide Web \cite{Fensel200301} confronts us with an information
overload: there are too much data and sources to be able to be found out those most relevant for us. Indeed, we have to make choices from
thousands of movies, millions of books, billions of web pages, and
so on. Evaluating all these alternatives by ourselves is not
feasible at all. As a consequence, an urgent problem is how to
automatically find out the relevant items for us. Internet search
engine \cite{Brin199801}, with the help of keyword-based queries, is
an essential tool in getting what we want from the web. However, the
search engine does not take into account personalization and returns
the same results for people with far different habits. In addition,
not all needs or tastes can be easily presented by keywords.
Comparatively, \emph{recommender system} \cite{Resnick199701}, which
adopts knowledge discovery techniques to provide personalized
recommendations, is now considered to be the most promising way to
efficiently filter out the overload information. Thus far,
recommender systems have successfully found applications in
e-commerce \cite{Schafer200101}, such as book recommendations in
\emph{Amazon.com} \cite{Linden200301}, movie recommendations in
\emph{Netflix.com} \cite{Bennett200701}, video recommendations in
\emph{TiVo.com} \cite{Ali200401}, and so on.

A recommender system is able to automatically provide personalized
recommendations based on the historical record of users' activities.
These activities are usually represented by the connections in a
user-item bipartite graph \cite{HuangZ200401,ZhouT200701}. So far, collaborative
filtering (CF) is the most successful technique in the design of
recommender systems \cite{Herlocker200401}, where a user will be
recommended items that people with similar tastes and preferences
liked in the past. Despite its success, the performance of CF is
strongly limited by the sparsity of data resulted from: (i) the huge
number of items is far beyond user's ability to evaluate even a
small fraction of them; (ii) users do not incentively wish to rate
the purchased/viewed items \cite{Resnick200001}.
Besides the fundamental user-item relations, some accessorial
information can be exploited to improve the algorithmic accuracy
\cite{Adomavicius200501}. User profiles, usually including age, sex,
nationality, job, etc., can be treated as prior known information to
filter out possibly irrelevant recommendations \cite{Kazienko200701},
however, the applications are mostly forbidden or strongly
restricted to respect personal privacy. Attribute-aware method
\cite{Tso200601} takes into account item attributes, which are defined
by domain experts. Yet it is limited to the attribute vocabulary,
and, on the other hand, attributes describe global properties of
items which are essentially not helpful to generate personalized
recommendations. In addition, content-based algorithms can provide
very accurate recommendations \cite{Pazzani200701}, however, they are
only effective if the items contain rich content information that
can be automatically extracted out, for example, these methods are
suitable for recommending books, articles and bookmarks, but not for videos, tracks or pictures.

Recently, the \emph{network theory} provides us a powerful and versatile tool to recognize and
analyze such relation-based complex systems where nodes represent individuals, and links denote the relations among them. Therefore, many social, biological
and technological and information systems can be represented as
complex networks. In addition, a vast amount of efforts has been
addressed in understanding the structure, evolution and dynamics of
complex networks \cite{Albert200201, Dorogovtsev200201,
Newman200301, Boccaletti200601, Costa200701}. However, the advent of \emph{Web 2.0} and its affiliated applications bring a new form of
user-centric paradigm which can not be fully described by
pre-existing models on neither unipartite or bipartite networks. One such
example is the user-driven emerging phenomenon, \emph{folksonomy}
\cite{Mathes200401}, which not only allows users to upload resources (bookmarks, photos, movies, publications, etc.) but also freely
assign them with user-defined words, so-called \emph{tags}.
Folksonomy requires no specific skills for user to participate in,
broadens the semantic relations among users and resources, and
eventually achieves its immediate success in a few years. Presently, a large number of such applications can be found online, such as
\emph{Del.icio.us}\footnote{http://del.icio.us/} (with tags of
bookmarks by users), \emph{MovieLens}\footnote{http://www.movielens.org/} (with ratings of movies by
users), \emph{CiteULike}\footnote{http://www.citeulike.com/} (with
tags of publications by users),
\emph{BibSonomy}\footnote{http://www.bibsonomy.org/} (with tags of
bookmarks and references by users),
\emph{Flickr}\footnote{http://www.flickr.com/} (with tags of images
by users), \emph{Last.fm}\footnote{http://www.last.fm/} (with tags
of music by users) etc. From the view of physics, all these online systems have performed similar statistical properties, e.g. Zipf's law like rank-frequency distribution \cite{Cattuto200701} and Heaps' laws growth phenomenon \cite{Cattuto200901}, between which the in-depth understanding are studied in recent works \cite{ZhangZK200801, LvLY201002}.
With the help of those platforms, users can not only store their own resources and manage them with
social tags, but also look into other users' collections to
find what they might be interested in by simply keeping track of the baskets with tags. Unlike traditional information management methods where words (or indices) are normally pre-defined by experts or
administrators, e.g. the library classification systems. A tagging
system allows users to create arbitrary tags that even do not exist
in dictionaries. Therefore, those user-defined tags can reflect user
behaviors and preferences with which users can easily make
acquaintance, collaborate and eventually form communities with
others who have similar interests \cite{Sen200601}.

\section{Overview of Tag-based Recommender Systems}

Nowadays, people are confronting huge amount of information and making much effort in searching relevant or interesting items. However, as discussed in previous section, it is impossible for individuals to filter metadata from various structures and massive number of sources, especially in a user-generated information era \cite{NovO200701}. The motivation of users' contribution is straightforward: they build their own data based on which they become further involved in web-based communications. Social tagging is becoming one of most popular tools in playing important rules among various social activities. Ding \emph{et al.} \cite{DingY200901} provided good overviews of social tagging systems with emphasis on both its social impact and ontology modeling.

As a consequence, social tags can be naturally considered as kind of additional yet useful resource for designing effective recommendation algorithms. Firstly, tags are freely associated by users, which can reflect their personalized preferences. Secondly, tags express the semantic relations among items, which can help evaluating the underlying item qualities. Thirdly, the co-occurrence properties of tags can be employed to build both user communities and item clusters, which be further made use of to find relevant yet interesting items for targeted individuals. Therefore, tags provide us a promising way to solve some stubborn problems in recommender systems, e.g. the \emph{cold-start} problem \cite{ZhangZK201003}.

Up to date, a remarkable amount of researches have discussed how to apply tags in the domain of recommender systems. Hotho \emph{et al.} \cite{Hotho200602} proposed a modified PageRank \cite{Brin199801} algorithm, namely \emph{FolkRank}, to rank tags in folksonomies with the assumption that important tags are given by important users, which is akin to HITS \cite{Kleinberg199901} algorithm in internet networks. The \emph{FolkRank} is then be adopted to recommend tags \cite{Jaschke200701}. In addition, due to the user-generated property, tags are considered to have high personalized information, hence can be used to design methods for both personalized searching \cite{XuS200801} and recommendation. A good overview of social bookmarking and its applications in recommender systems can be found in a recent Ph.D. thesis \cite{Bogers200901}. However, although tags are especially useful for both organizing and searching resources, there are also many studies arguing that not all tags can benefit recommendation \cite{BischoffK200801} because of the various limitations of tags, such as \emph{polysemy}, \emph{synonymy}, \emph{ambiguity} \cite{Mathes200401,WuH200601,SpiteriL200701,LeeSS200701}, etc. These flaws are also the side effects of the uncontrolled vocabulary, thus it remains some open issues in tagging systems: (i) singularity vs. plurality: e.g. the words \emph{cat} and \emph{cats} somehow have very similar meanings, however, refer to two different words in tagging systems; (ii) polysemy vs. synonymy: e.g. the word \emph{apple} may refer to a kind of fruit, while it can also indicate the well-known computer company, \emph{Apple Inc.}, as well as its products; on the other hand, the words \emph{mac}, \emph{macintosh}, and \emph{apple} all point to the products of \emph{Apple Inc.}, however, it fails again to uncover their underlying relations in tagging systems; (iii) different online tagging systems allow users to give different formats of tags, e.g. \emph{Del.icio.us} only allows words to be assigned, which subsequently results in compound words with various symbols (e.g. underline, dashline, colon, etc.), leading to an unlimited formats of metadata. Such freestyle tags additionally exemplify the explosion of observed datasets, hence interfere in the analyses of the structure and user behaviors in tagging systems. Recently, researches have devoted much effort to solve those issues. Firstly, clustering-based methods \cite{Shepitsen200801, Capocci200801} are proposed to alleviate the word reduction problem. Secondly, semantic methods are discussed to use ontology-based to organize tags and reveal the semantic relations among them \cite{MikaP200701, Kim200801}. Thirdly, dimension reduction and topic-based methods are put forward to discover the latent topics \cite{Symeonidis200801, WetzkerR200901}, and graph-based methods are proposed \cite{ZhangZK201001, LiuZ201001}  to solve the sparsity problem in large-scale datasets.

In the following, we firstly give the evaluation metrics measured in this survey. Secondly we summarize some of the most recent and prominent tag-aware recommendation algorithms, showing and discussing how they make use of the aforementioned representations to address the some unresolved issues in recommender systems. Basically, there are three kinds of recommendations in social tagging systems: (i) predicting friends to users; (ii) recommending items to users; (iii) pushing interesting topics (tags) to users. However, as mentioned above, the most urgent problem in information era is to filter irrelevant items for individuals, therefore, in this survey, we mainly discuss the second case, and introduce some related methods discussing (i) or (iii) if necessary. Finally, we conclude with comparison of the surveyed methods and outline some future challenges of tag-aware recommendation algorithms.

\section{Tag-Aware Recommendation Models}

Formally, a social tagging network consists of three different kind of communities: users, items and tags, which subsequently form an entry set of personalized folksonomy, \emph{personomy} ~\cite{Lipczak200801}, each follows the form $\mathbb{F}$=\{user, item, tag$_1$, tag$_2$, $\cdots$,
tag$_t$\}, where $t$ is the number of tags assigned to this item
by the very user. Correspondingly, in a recommender system, a full folksonomy can be considered in two ways: (i) to be consisted of three sets, respectively of users \emph{U} =
\{$U_{1}$,$U_{2}$,$\cdots$,$U_{n}$\}, items \emph{I} =
\{$I_{1}$,$I_{2}$,$\cdots$,$I_{m}$\}, and tags \emph{T} =
\{$T_{1}$,$T_{2}$,$\cdots$,$T_{r}$\}. Consequently, each binary relation can be described by a adjacent matrix, $A$, $A'$ and
$A''$ for user-item, item-tag and user-tag relations, respectively. If $U_i$ has collected $I_j$, we set $a_{ij}$ = 1, otherwise $a_{ij}$ = 0. Analogously, we set $a'_{jk}$ = 1 if $I_j$ has been assigned by the tag $T_k$, and $a'_{jk}$ = 0 otherwise. Furthermore, the users' preferences on tags can be represented by a adjacent matrix $A''$, where $a''_{ik}$ = 1 if $U_i$ has adopted $T_k$ and $a''_{ik}$ = 0 otherwise; (ii) a ternary \cite{Symeonidis200801, Rendle200901} or hypergraph \cite{Ghoshal200901, Zlatic200901, ZhangZK201002} based structure: only complete ternary relation is taken into account to be existed as a real link. That is to say, each relation of $(u,i,t)$, represented as an existing component $Y$ = 1, if it exists in a folksonomy $\mathbb{F}$, and $Y$ = 0 otherwise.

\subsection{Evaluation Metrics}
For a traditional recommender system, each data set, $E$, is randomly divided into two parts to test the performance of proposed algorithms: the training set, $E^P$, is treated as known information, while the testing set, $E^T$, is used for testing. In this survey, the training set always contains 90\% of entries,  and the remaining 10\% of entries, constitute the testing set. In addition, each division should guarantee $E^T \bigcap E^P = {\O}$ and  $E^T \bigcup E^P = E$ in order to make sure no redundant information is used. Furthermore, To give solid and comprehensive evaluation of the proposed algorithm, we consider metrics of both accuracy \cite{Gunawardana200901} and diversity \cite{ZhouT201001} to characterize the performance of recommendations.
\subsubsection{Metrics of Accuracy}
\begin{enumerate}
\item \emph{Ranking Score} ($RS$) \cite{ZhouT200701}.---
In the present case, for each entry in the testing set (i.e. a
user-item pair), $RS$ is defined as the rank of the item,
divided by the number of all uncollected items for the
corresponding user. Apparently, the less the $RS$, the more
accuracy the algorithm is. $\langle RS \rangle$ is given by
averaging over all entries in the testing set.

\item \emph{The area under the ROC curve} \cite{Hanley198201, Clauset200801}.---
In the present case, the area under the ROC curve, abbreviated by AUC, for a particular user is the probability that a randomly selected removed item for this user (i.e., an item in the testing set and being collected by this user) is given a higher score by our algorithm than a randomly selected uncollected item (i.e, an item irrelevant to this user in neither the training set nor the testing set). The AUC for the whole system is the average over all users. If all the scores are generated from an independent and identical distribution, $AUC\approx 0.5$. Therefore, the degree to which the AUC exceeds 0.5 indicates how much better the algorithm performs
than pure chance.

\item \emph{Recall} \cite{Herlocker200401}.---
Note that, the AUC takes into account the order of all uncollected
items, however, in the real applications, user might only care about
the recommended items, that is, the items with highest scores.
Therefore, as a complementary measure, recall is employed to
quantify the accuracy of recommended items, which is defined as:

\begin{equation}
  \emph{Recall}=\frac{1}{n}\sum^n_{i=1}N^i_r/N^i_p,
\end{equation}

where $N^i_p$ is the number of items collected by $U_i$ in the
testing set, and $N^i_r$ is the number of recovered items in the recommendations for $U_i$. We use the averaged recall instead of simply counting $N_r/N_p$ with $N_r=\sum_iN^i_r$ and
$N_p=\sum_iN^i_p$ since it is fair to give the same weight on every user in the algorithm evaluation. Assuming the length of
recommendation list, $L$, is fixed for every user, recall is very sensitive to $L$ and a larger $L$ generally gives a higher recall.
\end{enumerate}

\subsubsection{Metrics of Diversity}
\begin{enumerate}
\item \emph{Inter Diversity} ($InterD$) \cite{ZhouT200802,
ZhouT200701}.--- $InterD$ measures the differences of different users'
recommendation lists, thus can be understood as the inter-user
diversity. Denote $I^i_R$ the set of recommended items for user
$U_i$, then
\begin{equation}
    \emph{InterD} =
    \frac{2}{n(n-1)}\sum_{i\neq j}\left(1-\frac{|I^i_R\cap I^j_R|}{L}\right),
\end{equation}
where $L=|I^i_R|$  is the length of recommendation list. In average,
greater or less $InterD$ mean respectively greater or less
personalization of users' recommendation lists.

\item \emph{Inner Diversity} ($InnerD$) \cite{ZhouT200802}.---
$InnerD$ measures the differences of items within a user's
recommendation list, thus can be considered as the inner-user
diversity. It reads,
\begin{equation}
    \emph{InnerD} =
    1-\frac{2}{nL(L-1)}\sum^n_{i=1}\sum_{j\neq l,j,l\in I^i_R}S_{jl},
\end{equation}
where
$S_{jl}=\frac{|\Gamma_{I_j}\cap\Gamma_{I_l}|}{\sqrt{|\Gamma_{I_j}|\times
|\Gamma_{I_l}|}}$ is the cosine similarity between items $I_j$ and
$I_l$, where $\Gamma_{I_j}$ denotes the set of users having
collected object $I_j$. In average, greater or less $InnerD$
suggests respectively greater or less topic diversification of
users' recommendation lists.
\end{enumerate}
\subsection{Network-based Models}
Recently, there are a variety of attempts utilizing tagging information for recommendation from a perspective of graph theory,
Generally, a tag-based network can be viewed as a tripartite graph which consists of three integrated bipartite graphs \cite{ZhouT200701} or a hypergraph. Therefore, network-based methods are widely used to describe the tag-based graph. Up to date, bipartite graph has been largely applied to depict massive number of online applications. For example, users rate movies, customers comment books, individuals participate in online games, etc. In a typical bipartite graph, there are two mutually connected communities, which contrastively have no link within each community, shown in Fig. 1.

\begin{figure}
\centering
\centerline{\includegraphics[height=5.0cm, width=7.5cm]{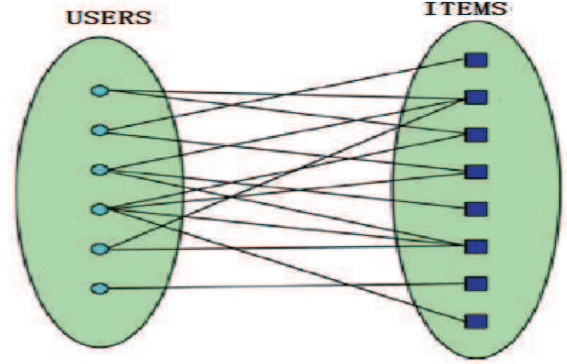}}
\caption{(Color online) Illustration of a user-item bipartite network  \cite{Shang201002} composed by 6 users and 8 items, in which only inter-community links are allowed.}
\end{figure}

Inspired by this elegant structure, two underlying network-based methods: Probability Spreading (ProbS) \cite{ZhouT200701, ZhouT201001} and Heat Spreading (HeatS) \cite{ZhangYC200701, ZhouT201001}, were proposed as a starting point to apply network theory in recommender systems.

\begin{figure*}
\centering
\centerline{\includegraphics[height=5.5cm, width=15.5cm]{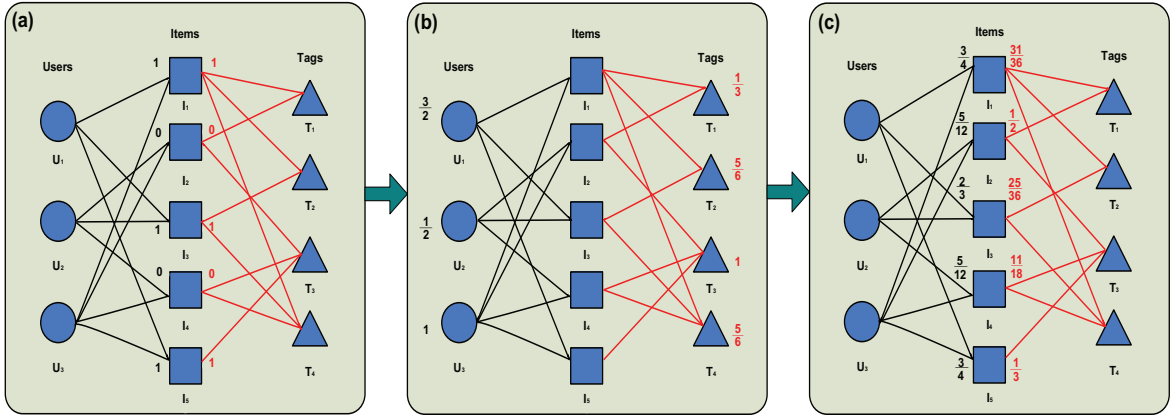}}
\caption{(Color online) Illustration of a user-item-tag tripartite graph consists of 3 users, 5 items and 4 tags, as well as the recommendation process described in \cite{ZhangZK201001}. The tripartite graph is decomposed to user-item (black links) and item-tag (red links) bipartite graphs connected by items. For the target user $U_1$, the scoring process works as: (a) firstly, highlight the items, $I_1$, $I_3$, $I_5$, collected by the target user $U_1$ and mark them with unit resource. That is to say: $f_{I_1}=f_{I_3}=f_{I_5}=1$, and $f_{I_2}=f_{I_4}=0$. (b) secondly, distribute the resources from items to their corresponding users and tags, respectively; e.g. $f_{U_3}=f_{I_1}*\frac{1}{2}+f_{I_2}*\frac{1}{2}+f_{I_5}*\frac{1}{2}=1*\frac{1}{2}+0+1*\frac{1}{2}=1$ and $f_{T_4}=f_{I_1}*\frac{1}{3}+f_{I_3}*\frac{1}{2}+f_{I_4}*\frac{1}{2}=1*\frac{1}{3}+1*\frac{1}{2}+0=\frac{5}{6}$; (c) finally, redistribute the resources from users and tags to their neighboring items. e.g. $f^p_{I_4}=f_{U_2}*\frac{1}{3}+f_{U_3}*\frac{1}{4}=\frac{1}{2}*\frac{1}{3}+1*\frac{1}{4}=\frac{5}{12}$ and $f^{pt}_{I_4}=f_{T_3}*\frac{1}{3}+f_{T_4}*\frac{1}{3}=1*\frac{1}{3}+\frac{5}{6}*\frac{1}{3}=\frac{11}{18}$.}
\end{figure*}

ProbS is also known as random walk (RW) in computer science and mass diffusion (MD) in physics.
Given a target user $U_i$, ProbS will generate final score of each item, $f_{j}$,
for her/him according to following rules:

Suppose that a kind of resource is initially located on
items. Each item averagely distributes its resource to all
neighboring users, and then each user redistributes the received
resource to all his/her collected items. The final resource vector
for the target user $U_i$, $\vec{f^p}$, after the two-step mass diffusion
is:
\begin{equation}
 f^p_j=\sum_{l=1}^n\sum_{s=1}^m\frac{a_{lj}a_{ls}a_{is}}{k(U_l)k(I_s)},
 \ j=1,2,\cdots,m,
\end{equation}
where $k(U_l)=\sum_{j=1}^ma_{lj}$ is the number of collected items
for user $U_l$, and $k(I_s)=\sum_{i=1}^na_{is}$ is the number of
neighboring users for item $I_s$.

Comparatively, HeatS works based on the reverse rules of ProbS. At each step,
each target will receive resources according to how active or popular it is, while ProbS distributes resources
based on its own activity or popularity. Thus, the final resource vector for the target user $U_i$, $\vec{f^h}$, after the two-step heat spreading is:
\begin{equation}
 f^h_j=\frac{1}{k(I_j)}\sum_{l=1}^n\sum_{s=1}^m\frac{a_{lj}a_{ls}a_{is}}{k(U_l)},
 \ j=1,2,\cdots,m,
\end{equation}

Therefore, HeatS will depress the score of popular items and is inclined to recommend the relatively cold ones, while ProbS will enhance the scoring ability of popular items.

Based on the aforementioned methods, a variety of algorithms have been proposed to add tags in order to generate better recommendation performance. Zhang \emph{et al.} \cite{ZhangZK201001} firstly proposed a tag-aware diffusion-based method, considering tags as additional information, which extended the resulting paradigm as reduced bipartite graphs, known as \emph{tripartite graph}. In such a graph, one kind of nodes (users, items or tags) plays as a centric role to bridge the remaining two. Fig. 2 shows an example of item-centric model. In such a graph, each item of a target user will respectively distribute to its neighboring users and tags, and then all the items in database will receive their resources from their neighboring nodes. Hence, the final resource for the target user $U_i$, $\vec{f^t}$, after two-step mass diffusion (see Fig. 2), will be integrated in a linear way:

\begin{equation}
f^t_j= \lambda  f^p_j + (1-\lambda)  f^{pt}_j,
\end{equation}
where $f^{pt}_j=\sum_{l=1}^r\sum_{s=1}^m\frac{a_{is}a'_{ls}a'_{lj}}{k'(I_s)k(T_l)}$ is the resource of item $j$ received from item-tag bipartite graph, $k(T_l)=\sum_{j=1}^ma'_{jl}$ is the number of neighboring
items for tag $T_l$, $k'(I_s)=\sum_{l=1}^ra'_{sl}$ is the number of
neighboring tags for item $I_s$, and $\lambda \in [0,1]$ is a tunable parameter to obtain the optimal performance. Table 1 shows the corresponding AUC results for three datasets: \emph{Del.icio.us}, \emph{MovieLens} and \emph{BibSonomy}, in which the AUC values are significantly improved by considering item-tag bipartite relation. In addition, \cite{ZhangZK201001} also experimentally demonstrated that the incorporation of tags can enhance the Recall results for various ranges of recommendation length. Besides the accuracy, \cite{ZhangZK201001} extensively showed that tags could also promote the recommendation diversification, hence enlarge the selection vision for users.

Recently, a variety of researchers have designed tag-aware algorithms by modifying the above model. Shang \emph{et al.} \cite {ShangMS201001} proposed a user-centric diffusion-based similarity, which considered users as the communication hubs to measure the coincidence among users, and found it could obtain more accurate recommendations. In addition, the tag usage frequency were measured as edge weight in user-item bipartite networks. Shang and Zhang \cite{ShangMS200901} directly regarded the frequency as weight and applied diffusion method to improve the recommendation accuracy. Wu and Zhang \cite{WuP201001} viewed the tag usage patten in a document vocabulary manner and applied the inverse document frequency (TF-IDF) model \cite{Salton198301} to calculate the weight for user-item relations. They found this weighting method could enhance the recommendation diversity. Furthermore, Zhang \emph{el al.} \cite{ZhangZK201003} took such tag usage frequency into account on the user-tag and then spread the tag-based preferences to all the corresponding tags' neighboring items. The numeric results showed it could significantly enhance the algorithmic accuracy for relatively inactive or new users, and it also found that different tag usage patterns might result in different algorithmic diversity: the more diverse topic of tags users like, the more diverse results the algorithm would generate. Consequently, two fundamental roles of tags \cite{ZhangZK201002, ZhangZK201101}, describing and retrieving items, were firstly found applications in recommender systems. Up to date, Liang \emph{et al.} \cite{LiangHZ201001} have noticed that the above methods decomposed the user-item-tag relationships into two bipartite graphs and made recommendations, which, to some extents, ignored the remaining one binary relation (e.g. user-tag for \cite{ZhangZK201001}, user-item for \cite{ZhangZK201003}). As a result, by further eliminating the noise of tags, they used the semantic meaning of tags to represent topic preferences of users and combined it with item preferences of users to measure user-based similarity. Subsequently, the hybrid similarity was used in a standard collaborative filtering framework to obtain better \emph{Recall} results in two datasets: \emph{Amazon.com} and \emph{CiteULike.com}. Similar measurements of user-based and item-based similarities were also widely applied by various researches \cite{Szomszor200701,Sutter200801}.

\begin{table*}
\centering \caption{Comparison of algorithmic accuracy, measured by
the AUC. \emph{Pure U-I} and \emph{Pure I-T} denote the pure
diffusions on user-item bipartite graphs and item-tag bipartite
graphs, respectively corresponding to $\lambda$=1 and $\lambda$=0.
The optimal values of $\lambda$ as well as the corresponding optima
of AUC are presented for comparison.}
\begin{tabular}{cccccc}  \hline \hline Data set & Pure U-I & Pure I-T & Optimum &
$\lambda_{opt} $
\\ \hline
\emph{Del.icio.us} &0.8098&0.8486&  0.8588 &   0.32 \\
\emph{MovieLens} &0.8065&0.8163  &0.8233 &  0.44\\
\emph{BibSonomy} &0.7374&0.7600 &0.7852  &   0.44 \\
\hline \hline
\end{tabular}
\end{table*}

\subsection{Tensor-based Models}
Recently, the tensor factorization (TF) \cite{KoldaTG200901} based method has attracted increasing attention to be applied in designing recommendation algorithms in social tagging systems \cite{XuY200601, Symeonidis200801, Symeonidis200901, Kantor201001, Rendle200901}. Generally, by using tensor, a ternary relation,  $\mathbb{A} =\{u,i,t\}$, can be represented as \cite{Symeonidis200901}

\begin{equation}
 a_{(u,i,t)} =
 \left\{
  \begin{array}{l}
 1,  if  (u,i,t) \subseteq \emph{Y},
 \\0, otherwise.
 \end{array}
\right.
\end{equation}

There are also other researches that define the missing values for empty triples in which the items have never been tagged, while the negative values are set for the triples in which the items are tagged in other tensors \cite{Rendle200901}. Fig. 3 shows the illustration of the above two definitions.

\begin{figure}
\centering
\centerline{\includegraphics[height=3.0cm, width=7.5cm]{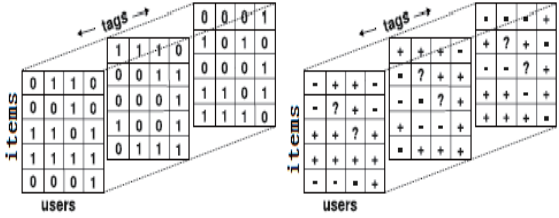}}
\caption{Illustration of tensor-based tag assignment ternary relation. Left panel: a visible tag assignment, $a_{(u,i,t)}$, is set 1, and 0 otherwise \cite{Symeonidis200901}. Right panel: $a_{(u,i,t)}$ is set negative as the triple of which the item is tagged in other existing triples rather than $a_{(u,i,t)}$. The missing values are given to other empty triples \cite{Rendle200901}.}
\end{figure}

For the purpose of recommendation, $Y$ can be represented by three low-rank \emph{feature metrics}, $\hat{U}$, $\hat{I}$, $\hat{T}$ and one \emph{core tensor}, $\hat{C}$, shown as

\begin{equation}
   \hat{Y} = \hat{C}_{\times u} \hat{U}_{\times i} \hat{I}_{\times t} \hat{T},
\end{equation}
where  the core tensor $\hat{C}$ and the feature matrices
$\hat{U}$ , $\hat{I}$ and $\hat{T}$ are the parameters to be
learned and $\times x$ is the tensor $x$-mode dimension multiplication factor between a tensor and a matrix \cite{Rendle200901}. In addition, the size of feature matrices are:
\begin{equation}
  \begin{array}{l}
 \hat{C} \subseteq \mathbb{R}^{k_U \times k_I \times k_T}, \hat{U} \subseteq \mathbb{R}^{|U| \times k_U},
 \\ \hat{I} \subseteq \mathbb{R}^{|I| \times k_I}, \hat{T} \subseteq \mathbb{R}^{|T| \times k_T},
 \end{array}
\end{equation}
where $k_U$, $k_I$, $k_T$ are the latent dimensions of the low-rank approximations for users, items and tags, respectively. Then, recommendations can be generated as
\begin{equation}
 \hat{y}_{(u,i,t)}= \sum_{\hat{u}}\sum_{\hat{i}}\sum_{\hat{t}}
 \hat{c}_{(\hat{u},\hat{i},\hat{t})} \cdot \hat{u}_{(u,\hat{u})}  \cdot \hat{i}_{(i,\hat{i})}  \cdot \hat{t}_{(t,\hat{t})},
\end{equation}
where the tilde denotes the feature dimensions and the hat indicates the elements of the feature matrices. Finally, the personalized recommendations list of items or tags  will be displayed to the target user in a descending order.

The tensor factorization is based on Singular Value Decomposition (SVD) \cite{Wall200301}, with which the ternary relation can be reduced to low dimensions, hence easier to be proceeded for recommendation. \cite{Symeonidis200801} used it corresponding to a TF model optimized for square-loss where all not observed values are learned as 0s. In further, \cite{Symeonidis200901} developed a unified framework to model the three types of entities. Then, the three-order tensor dimension decomposition was performed by combining Higher Order Singular Value Decomposition (HOSVD) \cite {LathauwerL200001} method and the Kernel-SVD \cite{Shawe200401, ChinTJ200601} smoothing technique on two real-world datasets: \emph{Last.fm} and \emph{BibSonomy}. The results showed improvements in Recall and Precision. \cite{Rendle200901} proposed a better learning approach for TF models, which optimized the model parameters for the AUC values. The
optimization of this model is related to Bayesian personalized ranking (BPR) proposed in \cite{Rendle201001}. They both tried to optimize over pairs of ranking constraints, where the former focused on
AUC optimization, and the latter optimized for pair classification.
\cite{Rendle200902} discussed the relationship between them in details.

\subsection{Topic-based Models}

Generally, the \emph{core} challenge of recommender systems is to estimate the likelihood between users and items. In the last two decades, many efforts have been devoted to build various models to measure such probabilities in information retrieval. Deerwester \emph{et al.} \cite{DeerwesterS199001} proposed Latent Semantic Analysis (LSA) to use a term-document matrix describing the occurrences of terms in documents. Normally, each element in the matrix is weighted by TF-IDF \cite{Salton198301} revealing the importance of the very term in its corresponding documents. In addition, Hofmann \cite{HofmannT199901} introduced the Probability Latent Semantic Analysis (PLSA) to improve recommendation quality for various settings by assuming a latent lower dimensional topic model as origin of observed co-occurrence distributions. Comparing with the standard LSA, PLSA is based on a mixture decomposition derived from a latent topic model which would statistically result in a more principled approach having a solid foundation.
Eq. 11 gives a formula way of PLSA
\begin{equation}
  \begin{array}{rcl}
   P(w,d)= \sum\limits_{z} P(z)P(d|z)P(w|z) \\
   = P(d) \sum\limits_{z} P(z|d)P(w|z),
   \end{array}
\end{equation}
where word $w$ and document $d$ are both generated from the latent topic $z$, which is chosen conditionally to the document according to $P(z|d)$, and a word is then generated from that topic according to $P(w|z)$. However, PLSA does not allocate the topic distribution for each document, which might potentially lose information of documents with multiple subjects. Therefore, recently, a more widely used model, Latent Dirichlet Allocation (LDA) \cite{BleiDM200301}, was proposed to overcome this issue by allowing multiple latent topics with a priori Dirichlet distribution, a conjugate prior of multinomial distribution, assigned to each single document. Besides, LDA assumes that the documents are represented as random mixtures over the latent topics, each of which is given by a distribution over words. For each document $d$ in collection $D$, LDA works as (see Fig. 4):

\begin{figure}
\centering
\centerline{\includegraphics[height=4.0cm, width=7.5cm]{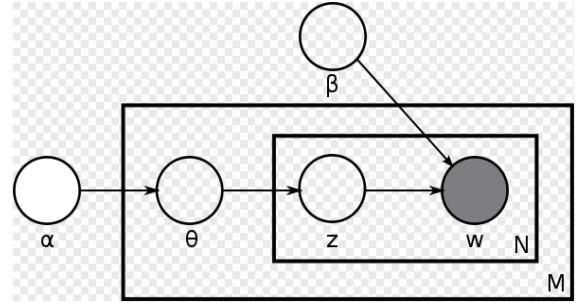}}
\caption{(Color online) Illustration of generative process for LDA model (from wikipedia.org), where $\alpha$, $\beta$, $\theta$ are parameters to be learned, $z$ is the latent topic variable, $w$ is observed variable of words, and the direction of arrows indicates the process flow. }
\end{figure}

(i) Choose $\theta_i$ from $Dir(\alpha)$, where $i$ runs over the document collection; (ii) For each word $w_{ij}$ in document $d_i$, choose a latent topic $z_{ij} \sim Multinomial(\theta_i)$ and then choose a word $w_{ij} \sim Multinomial(\beta_{z_{ij}})$. Finally, after learning the parameters by Gibbs sampling \cite{GelfandAE199001} or expectation-maximization (EM) algorithm \cite{DempsterAP197701}, the probability of the document collection can be given as

\begin{equation}
  \begin{array}{rcl}
   P(D|\alpha,\beta)&=& \prod\limits_{i} \int p(\theta_{i}|\alpha)*
\\&&\left(\prod\limits_{j}\sum\limits_{z_{ij}}p(z_{ij}|\theta_{i})p(w_{ij}|z_{ij},\beta)\right) d\theta_{i}
   \end{array}
\end{equation}

Recently, those topic-based models are applied in social tagging  systems for both tag and item recommendations. In \cite{WetzkerR200901, Umbrath200901}, they proposed a PLSA-based hybrid approach unifying user-item and item-tag co-occurrence to provide better item recommendations. In these two works, they measured the co-occurrence probabilities of user-item and item-tag by summing over the latent topic variables, and then maximized the likelihood of fused scenarios.

Comparatively, LDA is more widely used for tag recommendation. Xi \emph{et al.} \cite{SiX200901}  employed LDA for
eliciting topics from the words in documents, as well as the co-occurrence tags, where words and tags form independent
vocabulary spaces, and then recommended tags for target documents. Krestel \emph{et al.} \cite{KrestelR200901, KrestelR200902}, on the other hand, used LDA to extract hidden topics from the available tags of items and then recommended tags from these
latent topics. Bundschus \emph{et al.} \cite{BundschusM200901} integrated both user information and tag
information into LDA algorithm. Its generative process
extracted user specific latent topics using a Topic-Tag Model adding tags and User-Topic-Tag Model adding the user layer. It assumed that users had a multinomial distribution over topics, hence, the users' interests could be modeled by each tag assignment. Finally, they used two-step latent topic realizations (user-item based and tag-based topics) to provide personalized tag recommendations. In addition, Bundschus \emph{et al.} \cite{BundschusM200902} summarized different topic modeling approaches with respect to their ability to model annotations. Different from applying Bayesian rule to decompose the joint probability of item-tag and user-tag co-occupance, Harvey \emph{et al.} \cite{HarveyM201001} introduced a similar LDA-based approach for tag recommendation by decomposing
the joint probability of latent topics given the tag assignments.
Furthermore, Li \emph{et al.} \cite{LiD201001} combined LDA and GN community detection algorithm \cite{GirvanM200201, Leskovec201001} to observe the topic distributions of communities, as well as community evolving over time in social tagging systems. On this basis, they found that users in the same community tended to be interested in similar topics, which would shed some lights on recommendation for groups.

\section{Conclusions and Outlook}
In this survey, we summarized the progress of studies on tag-aware recommender systems (RS), emphasizing on the recent contributions by both statistical physicists and computer scientists in three aspects: (i) network-based methods; (ii) tensor-based methods; (iii) topic-based methods. Generally, there is no single method that can fully address all the problems existing in RS. Network-based and tensor-based methods can overcome the sparsity of large-scale data, hence can be used for designing efficient algorithms. However, they only focus on the network structure, while lack considerations of relations among tags. Comparatively, topic-based methods can distinguish tags into different topics, hence can produce more meaningful and understandable recommendations. However, since most of topic-based methods use machine learning to iteratively refine the results, they require high-efficient hardwares for computation, and thus consume more computation time. Similar problem lies in tensor-based methods for dimension reduction process. Therefore, a unified model might be considered to fully make use of their advantages and provide a more promising method in tag-aware recommender systems.

Nowadays, RS is not a new problem in information science, the advent of new Web2.0 paradigms bring versatile tools and information to help build better recommendation models by integrating traditional methods. Recently, the studies of complex networks would benefit tag-based algorithms, because the in-depth understanding of network structure, user behaviors and network dynamics can be used to design advanced tag-aware recommendation algorithms (e.g., making use of the information about hypergraph \cite{ZhangZK201002, WangJW201001} and tripartite graph \cite{Lambiotte200601, Cattuto200702} of social tagging networks to better predict underlying interests). On the other hand, tag-based algorithms can also help the trend detection \cite{Hotho200603} over time.

Although the studies of tag-aware recommender systems have achieved fruitful goals, there are still challenges, as well as some new directions remained to be solved (discovered) in future: (i) the complete hypergraph \cite{Zlatic200901, ZhangZK201002} should be well considered to fully address the integrity of tagging networks without decomposing any information and thus is a promising way to provide recommendations with better performance; (ii) most of current related researches emphasize on recommending single type of nodes, however, predicting the joint node pairs (e.g. item-tag pair \cite{PengJ201001}) comparatively lacks of study. The joint pair recommendation would provide more personalized preference, hence be a new application of tag-aware recommender systems; (iii) since the tags are freely assigned by users, which consequently results in much noise of added tags. Tag clustering \cite{Shepitsen200801, GemmellJ200801, RamageD200901} methods and anti-spam \cite{KoutrikaG200801} technique would be both promising ways to reduce the noise and help provide high-quality recommendations; (iv) the probability-based models are mainly used to provide tag recommendations in most recent researches, while how to well use them to benefit item recommendations is still an open challenge. In addition, those models would also help to prevent rumor spreading \cite{PittelB198701, KarpR200001} and trend detection \cite{KontostathisA200301}; (v) the multi-layered network \cite{MedhiD200001} consists of user social interactions, tag co-occurrence relations and user-item-tag ternary information can be considered to describe the hierarchical structure of social tagging systems, and thus the Social Network Analysis (SNA) \cite{ScottJ198801} and social influence \cite{Medo200901,Huang201001, CiminiG201101} based techniques can be used to provide more substantial recommendations, and social predictions \cite{HeymannP200801, LvLY201102} as well; (vi) most tagging platforms are dynamical systems and evolve over time \cite{DubinkoM200701, LiuC201101}, thus the study of human dynamics \cite{BarabasiAL200501} in analyzing the temporal behaviors and interests can provide real-time recommendations \cite{GantnerZ201001, XiangL201001}.

\section{Acknowledgements}
This work is partially supported by the Future and Emerging Technologies (FET) programs of the European Commission FP7-COSI-ICT (QLectives with grant No. 231200 and LiquidPub with grant No. 213360). Z.-K.Z. and T.Z. acknowledge the National Natural Science Foundation of China under the grant nos. 60973069 and 90924011, and the Sichuan Provincial Science and Technology Department (Grant No. 2010HH0002).

\bibliographystyle{unsrt}
\bibliography{bibi}

\biography{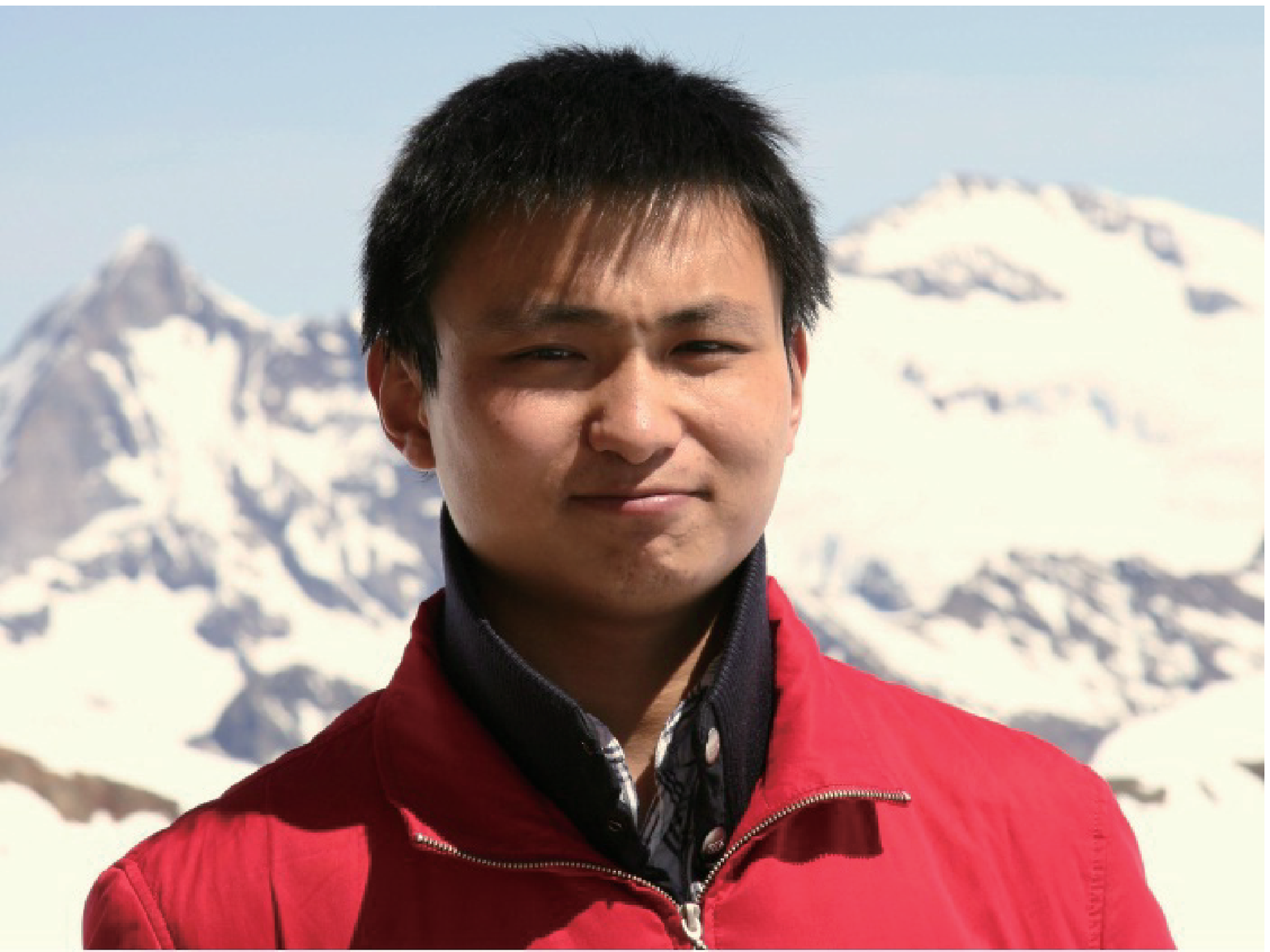}
{\bf Zi-Ke Zhang}
is a Ph.D. candidate of theoretical physics in University of Fribourg, Switzerland. His research interests include social tagging systems, recommender systems, complex networks, etc. He has published about 20 papers in the following journals: PLoS ONE, Physical Reviews, EPL, JSM and EPJB.

\biography{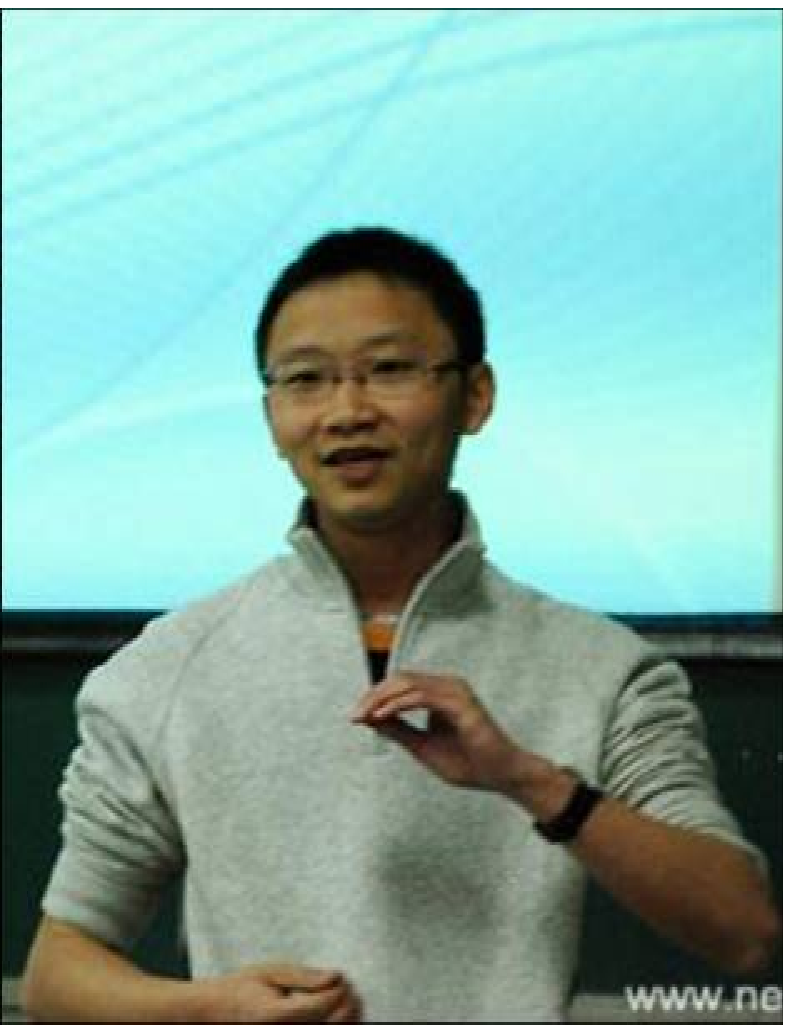}
{\bf Tao Zhou}
obtains Ph.D. majoring theoretical physics in University of Fribourg, Switzerland, as serves as a professor in the Web Sciences Center. His main research interests include complex networks, information physics, human dynamics, collective dynamics, and so on. He has published about 60 papers in the following five journals: PLoS ONE, Physical Reviews, EPL, NJP and PNAS. 

\biography{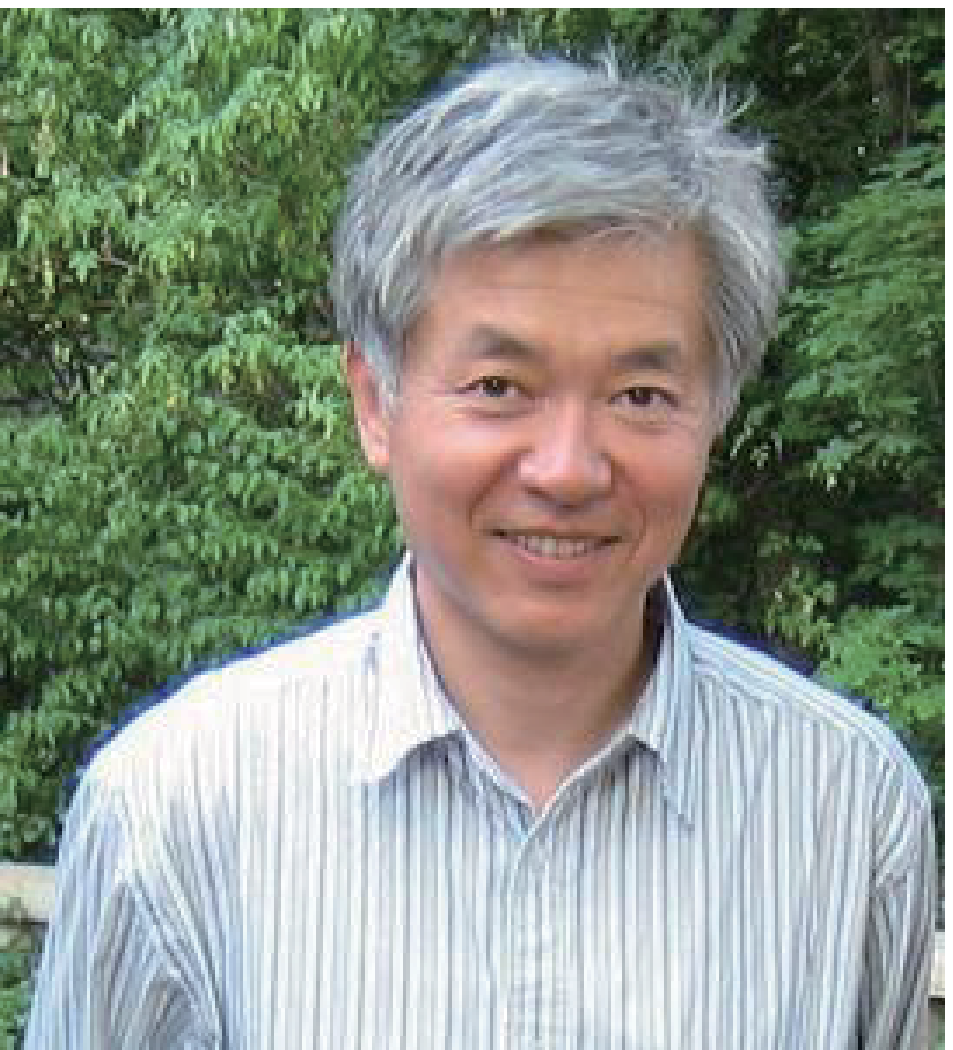}
{\bf Yi-Cheng Zhang}
is a professor of theoretical physics in University of Fribourg, Switzerland. He is the director of Institute for Information Economy in Hangzhou Normal University. His main research interests include complex systems, information physics, information economy, and so on. He has published about 130 papers in the following journals: Physics Report, PNAS, PLoS ONE, PRL, EPL, and NJP.

\end{multicols}
\end{CJK}
\end{document}